# An Ultra-fast Quantum Random Number Generation Scheme Based on Laser Phase Noise

Jie Yang[1,2], Mei Wu[2], Yichen Zhang[1], Jinlu Liu[2], Fan Fan[2], Yang Li[2], Wei Huang[2], Heng Wang[2], Yan Pan[2], Qi Su[3], Yiming Bian[1], Haoyuan Jiang[1], Jiayi Dou[1], Song Yu[1], Bingjie Xu[2,*], Bin Luo[1,†] and Hong Guo[4]

[1] State Key Laboratory of Information Photonics and Optical Communications, School of Electronic Engineering, Beijing University of Posts and Telecommunications, Beijing 100876, China
[2] Science and Technology on Security Communication Laboratory, Institute of Southwestern Communication, Chengdu 610041, China
[3] State Key Laboratory of Cryptology, Beijing 100878, China
[4] State Key Laboratory of Advanced Optical Communication Systems and Networks, School of Electronics and Center for Quantum Information Technology, Peking University, Beijing 100871, China

E-mail: *xbjpku@pku.edu.com, †luobin@bupt.edu.cn

**Abstract**

Based on the intrinsic random property of quantum mechanics, quantum random number generators allow for access of truly unpredictable random sequence and are now heading towards high performance and small miniaturization, among which a popular scheme is based on the laser phase noise. However, this scheme is generally limited in speed and implementation complexity, especially for chip integration. In this work, a general physical model based on wiener process for such schemes is introduced, which provides an approach to clearly explain the limitation on the generation rate and comprehensively optimize the system performance. We present an insight to exploit the potential bandwidth of the quantum entropy source that contains plentiful quantum randomness with a simple spectral filtering method and experimentally boost the bandwidth of the corresponding quantum entropy source to 20 GHz, based on which an ultra-fast generation rate of 218 Gbps is demonstrated, setting a new record for laser phase noise based schemes by one order of magnitude. Our proposal significantly enhances the ceiling speed of such schemes without requiring extra complex hardware, thus effectively benefits the corresponding chip integration with high performance and low implementation cost, which paves the way for its large-scale applications.

**Introduction**

Random numbers are an essential resource in a wide range of applications, such as statistical sampling, numerical simulations and cryptography. Pseudo random number generators based on computational algorithms provide an easy access to acquiring binary bit sequence that appears random and have been extensively employed in modern information systems. However, due to the deterministic and periodic nature, pseudo random number generators are not suitable for applications where true randomness is required, for instance, the cryptography [1] and quantum key distribution system [2, 3]. In contrast, quantum random number generators (QRNGs) extract randomness from quantum processes that can provide truly unpredictable and irreproducible random numbers [4, 5]. Based on the security assumptions for the system setup, existing QRNG schemes can be generally divided into three subcategories [4], i.e., the device-independent [6-11], semi-device-independent [12-21] and device dependent QRNG [22-47]. Device dependent QRNGs extract randomness from the implementation where physical devices for both the quantum state preparation and measurement are fully characterized and trusted, which to some extent comprises the security but makes it much more advantageous for practical applications [4, 5]. Over the past two decades, huge progress has been made in developing device dependent QRNGs with the generation rate enhanced from 1 Mbps to 100 Gbps.

To further improve the performance and practicality of device dependent QRNGs, it is desirable to exploit schemes beneficial for photonic chip integration, among which the laser phase noise (LPN) based QRNG schemes have attracted remarkable attentions [40-45]. The laser phase noise is a well understood quantum random phenomenon resulted from the inevitable spontaneous emission during lasing process which can be effectively distilled to generate random numbers. Till now, various such QRNG schemes have been proposed and demonstrated with off-the-shelf components or with photonic integrated unit [31-45], as shown in Table I. However, schemes by detecting the phase noise from a single laser requires unbalanced interferometer with long delay line and real-time feedback control, which increases the implementation complexity. As a comparison, schemes by detecting the phase noise between two independent lasers are preferred for chip integration, which removes the requirement for unbalanced interferometer and feedback control to effectively simplify the setup. Nevertheless, to achieve high performance on photonic chip, at least one laser needs to be operated in gain-switched mode, which requires extra hardware or equipment to generate high-speed driving pulses and the system performance is thus directly limited by the pulse generation speed [40, 43]. Therefore, how to realize high performance LPN based QRNG without need of high-speed driving pulses is of significant importance for practical application. Furthermore, in previous works, the randomness quantification and extraction are mainly based on the temporal and statistical analysis of the detection results, where detail analysis in frequency domain is incomplete, which is crucial for optimizing the bandwidth (BW) of quantum entropy source (QES) and improving the system performance.

In this paper, a general physical model applicable for QRNG schemes by detecting the laser phase noise is established and validated. The model is based on analytically illustrating the laser phase noise with wiener process and is effective for accurately predicting and analyzing the detection results both temporally and spectrally, which provides an approach to clearly explain the limitation on the generation rate and comprehensively optimize the system performance. Practically, a simple spectral filtering method is proposed based on detail frequency domain analysis and is verified experimentally through detecting the phase noise between two independent lasers operating in continuous-wave mode, which is significantly effective for optimizing and extracting the broad BW of the QES to boost the generation rate with no requirements for any extra high-speed driving pulses. Both the analog and digital filtering methods are designed and validated, based on which a real-time generation rate of 10 Gbps and an ultra-fast generation rate of 218 Gbps with off-line post-processing are experimentally demonstrated, respectively, setting a new record for LPN based QRNG schemes by one order of magnitude, as shown in Table I. It should be noted that with digital filtering, the pass band can be flexibly designed to match the spectrum of the practical detection result with a higher tolerance for environmental fluctuation. Compared with state-of-art for such QRNG schemes, our proposal provides an insight to exploit the potential bandwidth of the corresponding QES that contains plentiful quantum randomness which has not been practically explored in previous works and significantly enhances the ceiling speed of such schemes without requiring extra complex hardware, and thus benefits the corresponding chip integration with low implementation cost and high generation rate, as shown in Table I.

TABLE I. Comparison between our Results and state-of-art for high-speed QRNG base on Laser Phase Noise

| Reference | Laser Source | QES Bandwidth | Generation Rate | Need for real time phase control | Need for high speed pulse driver | Need for long delay line |
|---|---|---|---|---|---|---|
| Ref [33] | Single laser (CW) | 1 GHz | 6 Gbps (off-line) | Yes | No | Yes |
| Ref [35] | Single laser (GS) | GS at 5.8 GHz | 43 Gbps (off-line) | No | Yes | Yes |
| Ref [42] | Single laser (GS) | GS at 2.5 GHz | 10 Gbps (off-line) | No | Yes | Yes |
| Ref [40] | Dual lasers (CW + GS) | GS at 100 MHz | >1 Gbps (off-line) | No | Yes | No |
| Ref [43] | Dual lasers (GS + GS) | GS at 1 GHz | 8 Gbps (real-time) | No | Yes | No |
| Ref [38] | Dual lasers (CW+GS) | GS at 500 MHz | 2 Gbps (off-line) | No | Yes | No |
| Ref [38] | Dual lasers (CW+CW) | -- | 80 Mbps (off-line) | No | No | No |
| **This work** | **Dual lasers (CW+CW)** | **20 GHz (digital filtering) 1.5 GHz (analog filtering)** | **218 Gbps (off-line) 10 Gbps (real-time)** | **No** | **No** | **No** |

CW: continuous wave, GS: gain switched

## Results

*Physical model.* Typical experimental setups for QRNG schemes based on detecting the laser phase noise are shown in Fig. 1. In Fig. 1(a), the random phase shift of the optical signal from a single laser at two different times is distilled via an unbalanced Mach-Zehnder interferometer. The detection result at the photodiode with active feedback control of phase delay [33, 36] is generally given by $v(t) = A\sin\Delta\theta(t,T_d)$, where $T_d$ is the path delay between two arms, $\Delta\theta(t,T_d) = \theta(t+T_d) - \theta(t)$ is the random phase shift, and $A$ is a comprehensive coefficient that includes the amplitude of the optical signal, the attenuation of the optical path and the response of the photodiode, etc. In Fig. 1(b), the optical signals from two independent lasers interfere at a 50:50 beam-splitter (BS) to distill the phase noise and the detection result at the photodiode is generally given by $v(t) = A\cos(\Delta\omega t + \Delta\theta(t))$ where $\Delta\omega = 2\pi(f_1 - f_2) = 2\pi\Delta f$ and $\Delta\theta(t) = \theta_1(t) - \theta_2(t)$ are the difference of the angular frequencies and the phase noise fluctuations between the two lasers, respectively.

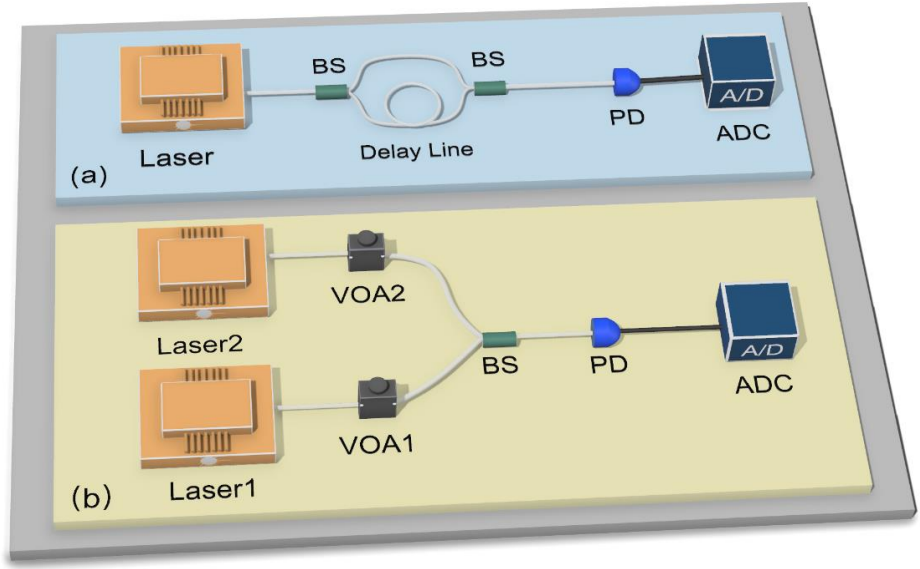

Figure 1. Schematic of the experimental setup for QRNG based on detecting the phase noise: (a) from a single laser and (b) between two independent lasers. PD: photodiode. $\tau_c$, $f$ and $\Delta v$: coherence time, central frequency and linewidth of the laser. BS: beam splitter. VOA: variable optical attenuator. $T_s$: sampling period.

For the scheme in Fig. 1(a), in previous works, $\Delta\theta(t,T_d)$ is directly modeled as a Gaussian random variable with variance $\langle\Delta\theta(t,T_d)\rangle^2 = 2T_d/\tau_c$ [48], where $\tau_c = 1/\pi\Delta v$ is the coherence time of the laser with linewidth $\Delta v$. Here, we present another physical model from the perspective of random process on how the phase fluctuation $\theta(t)$ evolves. With a sampling period $T_s$, the sampling results of analog-to-digital converter (ADC) can be expressed in a discretized form

$$v_1(nT_s) = A\sin\Delta\theta(nT_s,T_d) = A\sin\left[\theta(nT_s+T_d)-\theta(nT_s)\right] \quad (1)$$

where $\theta(nT_s+T_d)$ is essentially a delayed term of $\theta(nT_s)$ with $n=0,1,2,...$ Therefore once the random process $\theta(nT_s)$ is correctly modeled, the detection results $v_1(nT_s)$ can be obtained. Similarly, for the scheme in Fig. 1(b), the sampling results of ADC can be expressed as

$$\begin{aligned}v_2(nT_s) &= A\cos(\Delta\omega \cdot nT_s + \Delta\theta(nT_s)) \\ &= A\cos(2\pi\Delta f \cdot nT_s + \theta_1(nT_s) - \theta_2(nT_s))\end{aligned} \quad (2)$$

Therefore, as long as two independent random processes $\theta_1(nT_s)$ and $\theta_2(nT_s)$ are properly modeled, the detection result $v_2(nT_s)$ can be also obtained.

Fortunately, the random process of the laser phase fluctuation has been sufficiently investigated and can be modeled as a wiener process [48, 49]. As derived in [49], for a frequency-stabilized oscillator, by neglecting the amplitude fluctuations, it can be conducted that

$$\frac{d\theta(t)}{dt} = G(t)\cos(\theta + \omega_0 t) \tag{3}$$

where $\theta(t)$ is the phase term of the laser, $G(t)$ is the Gaussian input noise and $\omega_0$ is the angular frequency of the oscillator. Eq. (3) describes the nonlinear random process of the phase fluctuation and can be rewritten as

$$\frac{d\theta}{dt} = G(t)\cos(\theta + \omega_0 t) = F(t)e^{-i\phi} + F(t)^* e^{i\phi} \tag{4}$$

Here $F(t) = \frac{1}{2}Ge^{-i\omega_0 t}$ and $F(t)^* = \frac{1}{2}Ge^{i\omega_0 t}$ represent the random forces for the random process and can be described with $\langle F(t)F(t')^* \rangle \approx 2D_{FF^*}\delta(t-t')$, where $D_{FF^*}$ is a diffusion constant given $\omega_0 \Delta t \gg 1$ and can be evaluated by

$$\begin{aligned} 2D_{FF^*} &= (\Delta t)^{-1} \int_t^{t+\Delta t} ds \int_t^{t+\Delta t} ds' \langle F(s)F(s')^* \rangle \\ &= (2\pi\Delta t)^{-1} \int d\omega (G^2)_\omega \frac{\sin^2[(\omega-\omega_0)\Delta t/2]}{(\omega-\omega_0)^2} \\ &\approx \frac{1}{4}(G^2)_{\omega_0} \end{aligned} \tag{5}$$

The distribution function of the phase fluctuation $P(\theta,t)$ obeys the generalized Fokker-Planck equation

$$\frac{\partial P(\theta,t)}{\partial t} = \sum (-1)^n \frac{\partial}{\partial \theta^n}[D_n(\theta)P(\theta,t)] \tag{6}$$

where the $D_n(\theta)$ and the change in $\theta$ over the time interval $\Delta t$ are respectively defined by

$$D_n(\theta) = \langle[\theta(t+\Delta t) - \theta(t)]^n\rangle / \Delta t / n! \tag{7}$$

$$\Delta\theta = \theta(t+\Delta t) - \theta(t) = \int_{t}^{t+\Delta t}\left[F(s)e^{-i\theta(s)} + F(s)^{*}e^{i\theta(s)}\right]ds \tag{8}$$

With further derivation, it can be conducted that

$$\langle D_n \rangle = 0 \qquad n = 1, 3, 4, \ldots \tag{9}$$

$$2\langle D_2 \rangle = \frac{1}{\Delta t}\int_{t}^{t+\Delta t}ds\int_{t}^{t+\Delta t}ds'\langle F(s)F(s')^{*}\rangle + c.c.$$
$$= 2D_{FF^*} + c.c. = \frac{1}{2}(G^2)_{\omega_0} \tag{10}$$

Thus, Eq. (6) is simplified to the simple diffusion equation

$$\partial P/\partial t = \frac{1}{4}(G^2)_{\omega_0}\partial^2 P/\partial\theta^2 = D\partial^2 P/\partial\theta^2 \tag{11}$$

Finally, the conditional probability distribution of the phase fluctuation $P(\theta(t),t \mid \theta(0),0)$, which is the Green's-function solution of Eq. (11) can be obtained

$$P(\theta(t),t \mid \theta(0),0) = \frac{1}{\sqrt{4\pi Dt}} \times e^{-\frac{[\theta(t)-\theta(0)]^2}{4Dt}} \tag{12}$$

Eq. (12) demonstrates that the laser phase executes a simple Brownian motion and therefore can be modeled as a Wiener process, for which a more detailed demonstration is presented in Supplementary Note 1.

Specifically, for a laser with general Lorentzian spectrum, the phase fluctuation $\theta(t)$ can be modeled as a wiener process with $\theta(nT_s) - \theta((n-1)T_s) \sim N(0, 2\pi\Delta\nu T_s)$ [48], which properly provides an approach for the theoretical investigation of the LPN based QRNG.

***Simulation analysis and system performance optimization.*** As a result, one can theoretically simulate the detection results $v_1(nT_s)$ and $v_2(nT_s)$ in both time and frequency domain as shown in Box 1(a) and (b), respectively.

Box 1. Detail steps to simulate the detection results in both time and frequency domain for QRNG schemes based on phase noise (a) of a single laser and (b) between two independent lasers.

| (a) Simulation for the scheme of a single laser |
|---|
| **Step1**. Generate a Gaussian random variable $g(\Delta v, T_s) \sim N(0, 2\pi\Delta v T_s)$ |
| **Step2**. Generate the discretized wiener process $\theta(nT_s)$ by cumulatively summing up $g(\Delta v, T_s)$ |
| **Step3**. Given $T_d = mT_s$, generate $\theta(nT_s + mT_s)$ by delaying $\theta(nT_s)$ and calculate detection results $v_1(nT_s) = A\sin[\theta(nT_s + mT_s) - \theta(nT_s)]$ |
| **Step4**. Calculate the spectrum $V_1(k\Delta f)$ by performing Discrete Fourier Transform on $v_1(nT_s)$ |
| **(b) Simulation for the scheme of two independent lasers** |
| **Step1**. Generate two independent Gaussian random variables $g_1(\Delta v_1, T_s) \sim N(0, 2\pi\Delta v_1 T_s)$ and $g_2(\Delta v_2, T_s) \sim N(0, 2\pi\Delta v_2 T_s)$ |
| **Step2**. Generate the discretized wiener process $\theta_1(nT_s)$ and $\theta_2(nT_s)$ by cumulatively summing up $g_1(\Delta v_1, T_s)$ and $g_2(\Delta v_2, T_s)$ |
| **Step3**. Calculate detection results $v_2(nT_s) = A\cos[2\pi\Delta f \cdot nT_s + \theta_1(nT_s) - \theta_2(nT_s)]$ |
| **Step4**. Calculate the spectrum $V_2(k\Delta f)$ by performing Discrete Fourier Transform on $v_2(nT_s)$ |

Based on the proposed model and the corresponding simulation, both the temporal and spectral characteristic for the raw data of QRNG can be accurately predicted, which provides a numerical approach to evaluate the system performance quantitatively under various experimental parameters (e.g. $\tau_c, T_d, T_s$) and therefore a comprehensive optimization can be achieved. For example, for the QRNG scheme in Fig. 1(a), it is always advantageous to choose a laser with larger linewidth as shown in Supplementary Note 2, and one only needs to optimize $T_d$ and $T_s$ given $\tau_c$. The simulation results for different $T_d$ with fixed $\tau_c$ are performed in Fig. 2. When $T_d$ increases from 200ps to 1ns, the distribution uniformity of raw data increases which will enhance the extractable quantum randomness per sample $H_{\min}(M_{dis} | E)$ as explained specifically in ***methods*** [25, 38], meanwhile the flatness of spectrum that directly determines the BW of QES is remarkably compromised which in turn will limit the sampling rates $f_s$. Therefore, optimal $T_d$ and $f_s$ should be chosen to maximize $H_{\min}(M_{dis} | E) \times BW$, which can be numerically realized with our model. In previous works, one can only qualitatively conclude that one should increase the path delay and decrease the laser coherence time to enhance the distribution uniformity for better performance [31, 39], where various experiment tests need to be conducted to posteriorly verify the better experimental parameters. While the proposed model in this work in essence poses a more accurate illustration for the evolution of the random phase fluctuation, based on which the numerical optimization can be quantitatively realized in priori.

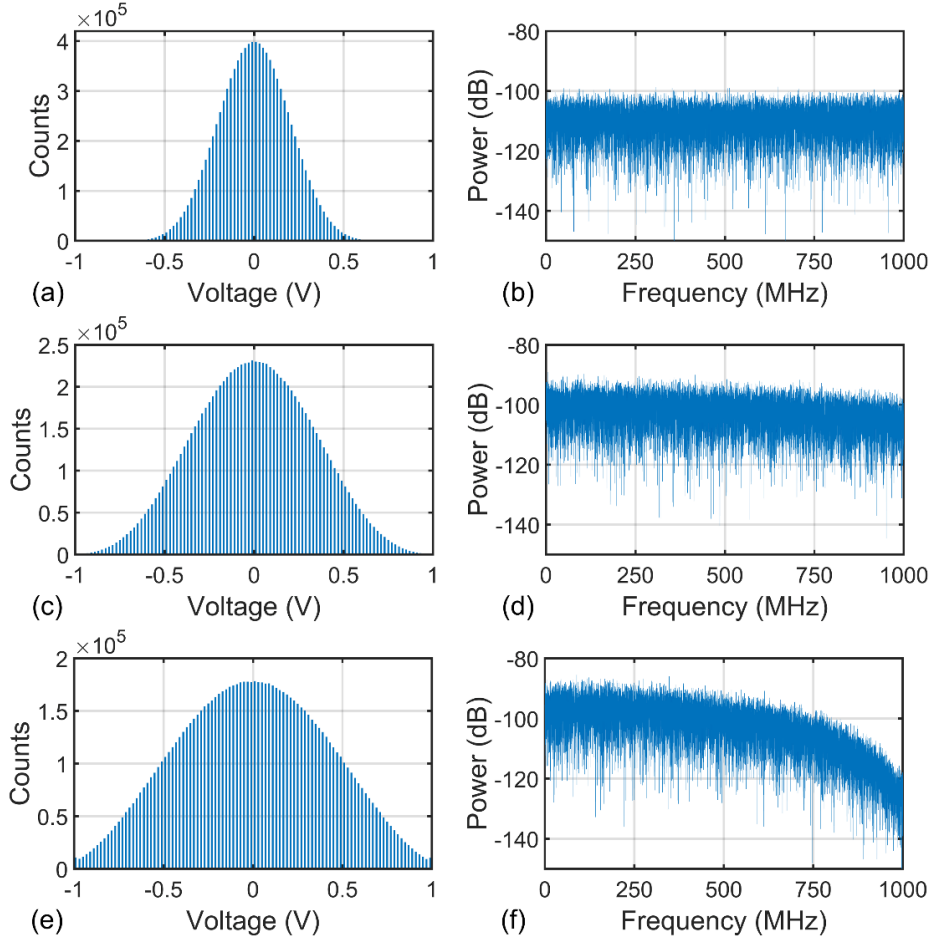

Figure 2. Simulation for histograms and power spectrums of the detection result calculated respectively from $10^7$ samples for (a, b) $T_d = 200 ps$, (c, d) $T_d = 600 ps$ and (e, f) $T_d = 1ns$. The laser coherence time is fixed at $\tau_c = 10ns$, sampling frequency is set as $f_s = 5GSa/s$ ($T_s = 200 ps$) and the comprehensive coefficient is set as $A = 1V$.

Though QRNG schemes based on detecting the phase noise from a single laser in Fig. 1(a) have been extensively demonstrated, an unbalanced interferometer with delay line and an extra real-time feedback control is generally required, which increases the implementation complexity and is not suitable for chip integration

[36]. To overcome this drawback, QRNG schemes by detecting the phase noise between two independent lasers are further proposed and studied [38], where the system setup is simplified effectively. But in previous works [38, 39], when both lasers operating in continuous-wave mode, the sampling rate is severely restricted and the generation rate is significantly limited. Thus, to achieve even higher generation rate, at least one laser is to be operated in gain-switched mode to emit high-speed optical pulse trains, which requires extra complex hardware or equipment to generate high-speed driving pulses [34, 35, 40, 42, 43]. Therefore, how to realize high performance LPN based QRNG without need of high-speed laser pulses is of great importance for practical application. In the following, the detection of the phase noise between two independent lasers operating in continuous-wave mode, as shown in Fig. 1(b), is analyzed based on the proposed model, through which the restrictions on generation rate are clearly explained and the corresponding solution is introduced.

The simulation for different laser linewidths $\Delta v_1$ and $\Delta v_2$ with beating frequency $\Delta f = 1.9 GHz$ is performed in Fig. 3. From Fig. 3 (a, c, e), the detection result follows an arcsine distribution. Note that here the phase noise is the difference between two independent wiener processes and thus naturally approximates a uniform distribution, which is a significant advantage over the scheme based on a single laser, where the phase noise essentially obeys Gaussian distribution. The very interesting phenomena is revealed through the spectrum characteristic. From Fig. 3 (b, d, f), it is shown that when the linewidth is 100 kHz and 1 MHz, an obvious peak power can be observed in both spectrums and the -3 dB BW of the corresponding QES are 0.14 MHz and 1.73 MHz, respectively. While when the linewidth is increased to 100 MHz, the peak power becomes much less significant and the -3 dB BW of QES increases to 200.43 MHz, which allows for a much higher sampling rate. Note that the sampling rate for a specific QRNG is directly limited by the BW of QES to avoid oversampling that might lead to correlations between ADC samples of raw data, which is further explained detailly in Supplementary Note 3 based on our model. Thus, the performance of the QRNG scheme in Fig. 1(b) is mainly limited by the BW of QES [38, 39]. To lift the restriction, one may employ lasers with large linewidth even up to GHz, which, however, cannot be fulfilled with the most commonly used and cost-effective DFB laser. Moreover, it should be noticed that, increasing the laser linewidth or employing more unstable sources might result in the drifts of the laser beating [39], which enhances the difficulty to achieve the experimental stability.

Hence, how to enhance the BW of QES for such QRNG scheme with narrow linewidth DFB lasers is an interesting question.

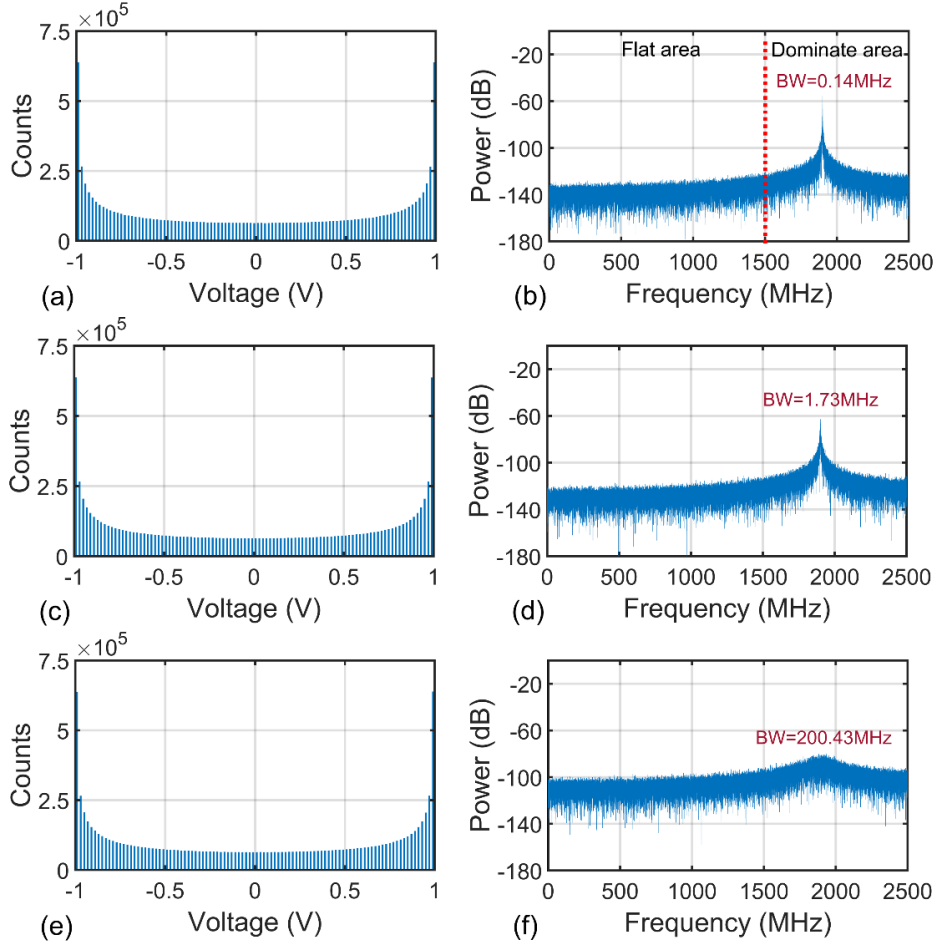

Figure 3. Simulation for histograms and power spectrums of the detection result calculated respectively from $10^7$ samples for (a, b) $\Delta v_1 = \Delta v_2 = 100kHz$, (c, d) $\Delta v_1 = \Delta v_2 = 1MHz$ and (e, f) $\Delta v_1 = \Delta v_2 = 100MHz$. The beating frequency is fixed at $\Delta f = 1.9GHz$, the sampling frequency is set as $f_s = 5GSa/s$ ($T_s = 200ps$) and the comprehensive coefficient is set as $A = 1V$.

To solve the problem, through further analysis of the signal spectrum, a simple and practical spectral filtering method is proposed, based on which general continuous-wave lasers with narrow linewidth can be employed to realize high speed quantum random number generation with high stability and thus the trade-off between the system performance and realization difficulty is achieved. Take the scenario with linewidth 100 kHz in Fig. 3 (b) for instance, the spectrum can be approximately divided into the dominate area, which is 1.5~2.5 GHz, and the flat area, which is 0~1.5 GHz. In the dominate area, a significant broadened peak of about -60 dB in power centered at 1.9 GHz can be observed, which theoretically dominates the power of the signal. While in the flat area, the power level drastically decreases to about -130 dB but with a bright flatness. This spectrum shape can be in principle explained as follows. On one hand, the signal $v_2(nT_s)$ in Eq. (2) is essentially the cosine function of the sum of two terms. The first term is $2\pi\Delta f \times nT_s$, which is periodic with the beating frequency $\Delta f$ and theoretically corresponds to a peak at $\Delta f$ in frequency domain. The second term is $\theta_1(nT_s) - \theta_2(nT_s)$, which is a random variable similar to white noise and theoretically spreads uniformly over the whole frequency domain. On the other hand, $\theta_1(nT_s) - \theta_2(nT_s)$ can be treated as a random phase jitter [38] that does not change the cosine feature of $v_2(nT_s)$, which indicates that the corresponding spectrum is in principle dominated by the term $2\pi\Delta f \cdot nT_s$. Therefore, the power spectrum of $v_2(nT_s)$ should be a peak at frequency $\Delta f$ and gradually decay to a stable level in other range with flatness, which is in accordance with our simulation results.

Based on the above analysis in frequency domain, it is theoretically feasible to eliminate the periodicity of the sampled data by spectrally filtering the dominate area that contains the broadened peak and thus the flat area that contains the quantum randomness can be effectively extracted, where, to our best knowledge, for the first time the division of the spectral areas in such QRNG scheme is introduced, inspiring the significant boost of the available bandwidth of QES to explore the potential randomness. As a result, the sampling rate and the generation rate can be significantly enhanced. Take the scenario under a linewidth of 100 kHz shown in Fig. 3 (b) for instance, it is inferred that by employing a low-pass filter with a cut-off frequency of 1.5 GHz, the flat area of 0~1.5 GHz can be distilled and the bandwidth of QES is drastically boosted by over $10^4$ times compared with original BW of 0.14 MHz. Notably, with proper design of the beating frequency and adequately fast detection and acquisition, the bandwidth of the QES is potential to reach even over 100 GHz.

***Experimental Demonstration.*** The experiment to verify our proposed model is performed with the setup shown in Fig. 1(b). Two DFB lasers with central wavelengths of 1550 nm and typical linewidths of 100 kHz in continuous-wave mode are employed as the optical source. Optical powers from both lasers at the photodiode are fixed at $P = 0.15mW$ by controlling the VOAs. Especially, two temperature controllers with an accuracy of 0.01 ℃ are employed to precisely shift the beating frequency, which is set around $\Delta f = 1.9GHz$ in the experiment. The output of the BS is detected by a photodiode with bandwidth 2GHz and then a high-performance digital oscilloscope (Agilent DSAZ254A, 25GHz bandwidth, 8-bit resolution ADC) is employed as the ADC to acquire the raw data. To effectively restore the signal feature and verify the proposed model, the sampling rate is set to $f_s = 20GSa/s$. By turning on and off the laser sources, the raw data and the electrical noise can be acquired and analyzed, respectively. As a comparison, the corresponding simulations based on the proposed model for the same setup is performed, where the comprehensive coefficient is experimentally measured to be $A = 0.87V$. The result is shown in Fig. 4.

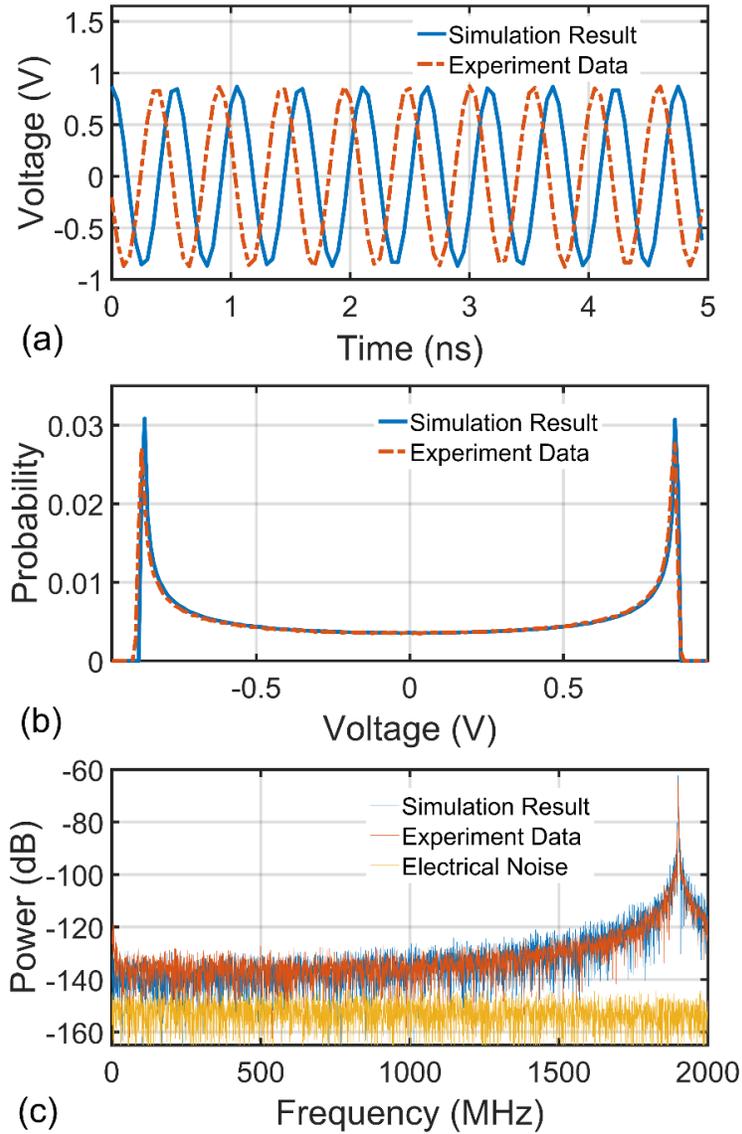

Figure 4．The agreement between the experimental data and the simulation results with respect to the (a) Temporal waveform, (b) Histogram, (c) Power spectrum. The linewidth of each laser is $\Delta v_1 = \Delta v_2 = 100 kHz$, beating frequency is $\Delta f = 1.9 GHz$, sampling rate is $f_s = 20 GSa/s$ and comprehensive coefficient is $A = 0.87V$. The electrical noise is added in the simulation results.

It is observed in Fig. 4 that significant agreements between the simulation results based on our model and the experimental data are achieved with respect to the temporal waveform, the histogram and the power spectrum, which sufficiently verifies the validation of the proposed model. Especially, in Fig. 4(c), the electrical noise is also presented to verify the quantum to classical noise ratio and a clearance of about 10dB between the raw data and the electrical noise is observed within the flat area, indicating that the quantum signal dominates the power in the corresponding frequency range and the random number generation based on the spectral filtering method should be feasible. Similar experiment results have also been achieved under other system parameters where remarkable agreements are still obtained.

To extract the frequency band of the flat area with high quantum to classical noise ratio, both the analog and digital filtering methods are designed and implemented, as shown in Fig. 5 (a) and (b) respectively, where one can either directly employ the off-the-shelf products to realize the analog filtering or integrate the digital filtering in the field programmable gate array (FPGA) as part of the post-processing, thus no extra hardware are required.

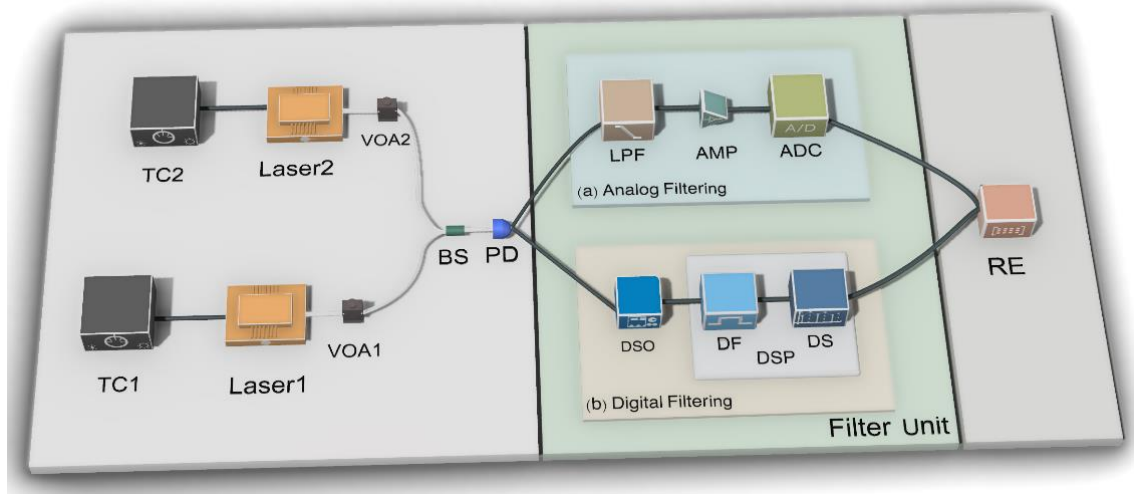

Figure 5. Schematic of the experimental setup for QRNG based on (a) analog filtering method and (b) digital filtering method. LPF: low pass filter. AMP: amplifier. RE: randomness extractor. DSO: digital storage oscilloscope. DSP: digital signal processing. DF: digital filter. DS: down sampling.

In the experiment of the analog filtering method in Fig. 5(a), the output signal of the BS is detected by a photodiode with bandwidth 2 GHz and the beating frequency is set around $\Delta f = 1.9 GHz$. Then an analog low pass filter (LPF) with a cut-off frequency of 1.5 GHz is employed to extract the frequency band of the flat area and an amplifier is employed to amplify the signal, which is processed by an ADC and a randomness extractor (RE) to generate the final random bits in real time. The experimental results are shown in Fig. 6.

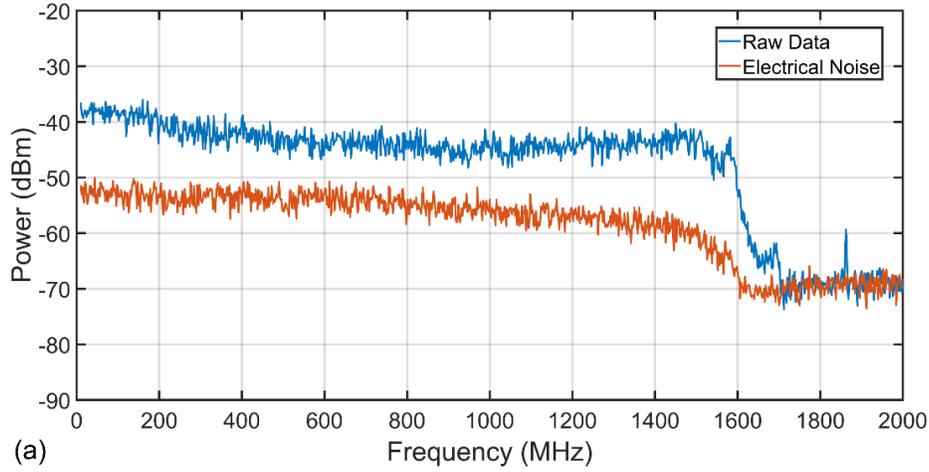
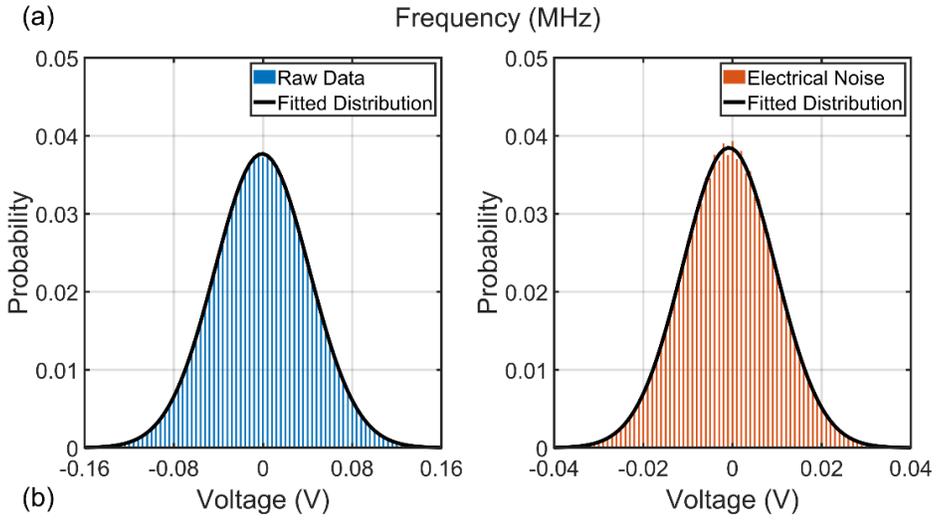

Figure 6. The experimental results of the QRNG based on the analog filtering method. (a) Power spectrum of the raw data and the electrical noise. (b) Histogram of the ADC sample data.

From Fig. 6 (a), it is shown that by employing the analog LPF, the frequency band of 0~1.5GHz is successfully extracted, where the raw data and the electrical noise both show a flat spectrum with a clearance of about 10dB. Especially, the histograms of the raw data and the electrical noise acquired by the ADC are illustrated in Fig. 6 (b), respectively, which both indicate a Gaussian distribution and fit significantly well with theoretical distribution curves and the conditional min-entropy is calculated to be 11.0407. With the real-time randomness extraction presented in *methods*, a real-time generation rate of 10 Gbps is achieved.

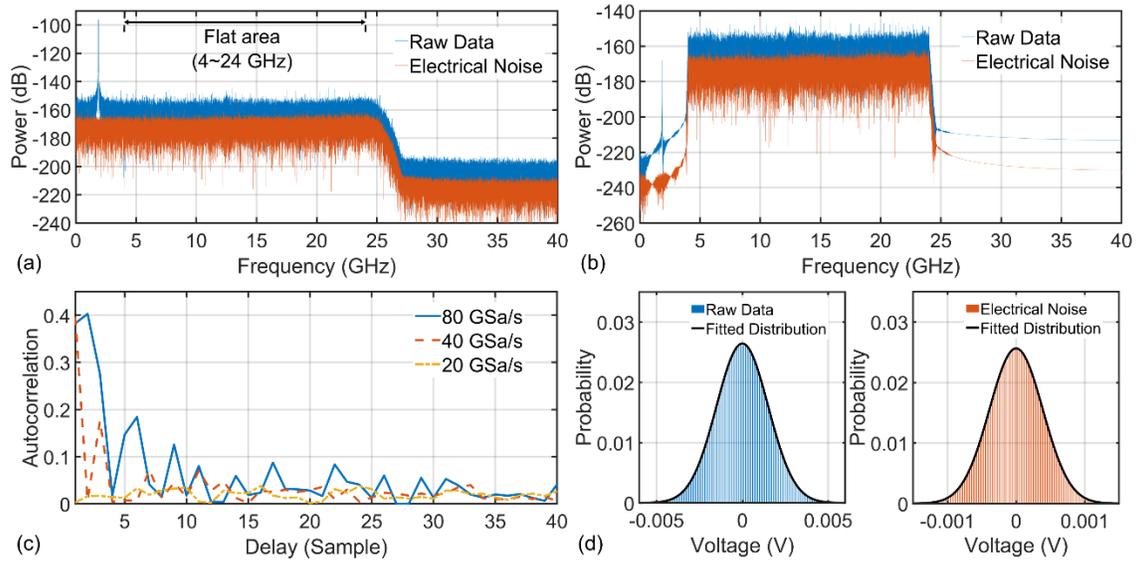

Figure 7. The experimental results of the QRNG based on digital filtering method. (a) Power spectrum of the raw data (blue curve) and the electrical noise (red curve) before DF. (b) Power spectrum of the raw data (blue curve) and the electrical noise (red curve) after DF. (c) Autocorrelation of the raw data after DF for sampling rate of 80 GSa/s (solid blue curve), 40 GSa/s (dashed red curve) and 20 GSa/s (dash-dotted yellow curve), respectively. (d) Histograms of the raw data (blue bar) and the electrical noise (red bar) after DF and DS. DF: digital filter. DS: down sampling.

In the experiment of the digital filtering method in Fig. 5(b), a high-speed photodiode with bandwidth 50GHz is employed to optimize the system performance, allowing for an ultra-fast generation rate. The beating frequency is set around $\Delta f = 1.9 GHz$. The output of the photodiode is digitized via the digital storage oscilloscope (DSO) at a sampling rate of $f_s = 80 GSa/s$, after which the sampled data is processed in the digital signal processing (DSP) stage. In the DSP stage, based on the spectrum of the raw data, the frequency band of 4~24 GHz is determined as the flat area as shown in Fig. 7(a). Then, a 10-th order elliptic band-pass digital filter is designed and applied to extract the frequency band of flat area, achieving a QES bandwidth of 20 GHz as shown in Fig. 7 (b). After the digital filter, to lower the autocorrelation between samples, the filtered raw data is down sampled to 40 GSa/s or 20 GSa/s, respectively, and autocorrelations of the filtered raw data with different sampling rates are calculated, as shown in Fig. 7 (c). For the balance of the generation rate and randomness, the filtered raw data is down sampled to 40 GSa/s, which is exactly twice the bandwidth of filtered

QES [28]. The histograms of the data after DSP are depicted in Fig. 7(d), where the raw data and electrical noise both obey Gaussian distribution with a conditional min-entropy of 5.546. Finally, the randomness extraction is implemented and an ultra-fast generation rate of 218 Gbps is achieved. Detail analysis results of the DSP stage and randomness extraction are shown in *methods*.

Based on the experimental results, it is expected that by employing detection and acquisition with even higher speed, the BW of QES can be further enhanced and effectively extracted with the digital filtering method to realize the even higher generation rate. Especially, in practical experiments the shift of the beating frequency is inevitable due to environmental fluctuation, which results in the variation of the target flat area. In this scenario, the pass band of the digital filter can be flexibly designed to match the practical spectrum of the acquired data. Thus, compared with the analog filtering method, where the pass band is fixed with the employed analog filter, the digital filtering method allows for a higher tolerance of the environmental fluctuation and is of significant feasibility for other QRNG schemes [28].

**Discussion**

In this work, we have proposed a general physical model based on wiener process for LPN based QRNG schemes to clearly explain the limitation on the generation rate and comprehensively optimize the system performance through detail temporal and spectral analysis. Especially, for the first time the division of the spectrum in such QRNG schemes is introduced and explained, based on which the available bandwidth of QES is boosted up to 20 GHz and an ultra-fast generation rate of 218 Gbps is realized experimentally with a practical spectral filtering method. In particular, with our proposal, it is expected to achieve a QES bandwidth of over 100 GHz with adequately fast detection, which might not be the upper bound and leaves to be further explored.

For future study, it is expected to improve the security analysis framework for our proposal. For instance, it will be a valuable work to develop a theoretical model that accounts for the non i.i.d feature of raw data and the quantum side information [27] under composable security framework.

Our work significantly enhances the ceiling speed of LPN based QRNG schemes without requirements for long optical delay-line or extra complex hardware, which particularly benefits the forthcoming chip integration with low implementation cost and high generation rate. Furthermore, our proposal is expected to be suitable for practical applications that require huge amounts of random numbers with fast generation rate and compact design. For instance, one promising application is the integrated photonic QKD systems [44, 50, 51], where our proposal can be employed to generate random numbers for the quantum state preparation, detection and post-processing, base on which we believe our proposal effectively contributes to the maturation and miniaturization of the QKD system and poses part of the solution for the deployment of the future global quantum secure communication network. Besides, our work can also be applied in commercial fields to achieve high performance and security, such as the large-scale cloud platform, the next generation internet of things, etc.

**Methods**

*Conditional min-entropy.* Since in practical experiments, the classical electrical noise might be known or even controlled by an adversary, thus to further distill the secure quantum randomness from the raw data of the measurement, the conditional min-entropy proposed in Ref [25] for each ADC sample is calculated as follows.

In our experiments, the measured total signal $M$ can be modeled as $M = Q + E$. Here $Q$ and $E$ respectively represent the quantum noise signal with probability density function (PDF) $p_Q$ and the classical noise signal with PDF $p_E$, which are generally assumed to be statistically independent. As observed in the experiments, $Q$ and $E$ both obey the Gaussian distribution with zero mean and variance $\sigma_Q^2$ and $\sigma_E^2$, thus the PDF of $M$ can be expressed as

$$p_M(m) = \frac{1}{\sqrt{2\pi}\sigma_M} \exp\left(-\frac{m^2}{2\sigma_M^2}\right) \tag{13}$$

for $m \in M$ where the measurement variance $\sigma_M^2 = \sigma_Q^2 + \sigma_E^2$. Therefore, the conditional PDF between the measured signal $M$ and the classical noise $E$ is given by

$$\begin{aligned} p_{M|E}(m|e) &= \frac{1}{\sqrt{2\pi(\sigma_M^2 - \sigma_E^2)}} \exp\left(-\frac{(m-e)^2}{2(\sigma_M^2 - \sigma_E^2)}\right) \\ &= \frac{1}{\sqrt{2\pi}\sigma_Q} \exp\left(-\frac{(m-e)^2}{2\sigma_Q^2}\right) \end{aligned} \tag{14}$$

Then the sampling is performed with through an $n$-bit ADC with dynamic range $[-R+\delta/2, R-3\delta/2]$, where $\delta = R/2^{n-1}$ represents the sampling bin width. Correspondingly, the discretized conditional probability distribution is

$$P_{M_{dis}|E}(m_i | e)$$
$$= \begin{cases} \int_{-\infty}^{-R+\delta/2} p_{M|E}(m|e) dm, & i = i_{\min}, \\ \int_{m_i-\delta/2}^{m_i+\delta/2} p_{M|E}(m|e) dm, & i_{\min} < i < i_{\max}, \\ \int_{R-3\delta/2}^{\infty} p_{M|E}(m|e) dm, & i = i_{\max} \end{cases} \quad (15)$$

where $m_i = \delta \times i$ and $i$ is a integer $\in \{-2^{n-1}, ..., 2^{n-1}-1\}$ with $i_{\min}$ and $i_{\max}$ represent the first and last bin to account for the saturation, respectively.

Therefore, the worst-case min-entropy conditioned on the classical noise $E$ for the discretized measured signal is given by [25]

$$H_{\min}(M_{dis} | E) = -\log_2 \left[ \max_{e \in R} \max_{m_i \in M_{dis}} P_{M_{dis}|E}(m_i | e) \right] \quad (16)$$

By performing integration on Eq. (15), the maximization for Eq. (16) can be conducted

$$\max_{e \in [e_{\min}, e_{\max}]} \max_{m_i \in M_{dis}} P_{M_{dis}|E}(m_i | e)$$
$$= \max \begin{cases} \frac{1}{2}\left[1 - erf\left(\frac{e_{\min} + R - \delta/2}{\sqrt{2}}\right)\right], \\ erf\left(\frac{\delta}{2\sqrt{2}}\right), \\ \frac{1}{2}\left[erf\left(\frac{e_{\max} - R - 3\delta/2}{\sqrt{2}}\right) + 1\right] \end{cases} \quad (17)$$

where $erf(x) = 2/\sqrt{\pi} \int_0^x e^{-t^2} dt$ is the error function and the classical noise is bounded in $[e_{\min}, e_{\max}]$ with generally $e_{\min/\max} = \pm 5\sigma_E$ or $e_{\min/\max} = \pm 10\sigma_E$ for a practical scenario. With Eq. (17), the secure randomness of each sample conditioned that the classical noise is fully known with arbitrary precision to the eavesdropper can be calculated.

***Real-time randomness extraction for the analog filtering.*** In the experiment of the analog filtering method in Fig. 5(a), with the ADC sampled data, it is calculated that the variance of raw data and electrical noise are $\sigma_M^2 = 1.792 \times 10^{-3} V^2$ and $\sigma_E^2 = 1.075 \times 10^{-4} V^2$, respectively, which infers that the quantum noise signal theoretically obeys Gaussian distribution with variance $\sigma_Q^2 = \sigma_M^2 - \sigma_E^2 = 1.684 \times 10^{-3} V^2$. The ADC employed is with a sampling rage of $2R = 0.8V$, a digitization resolution of $n = 14$ and a sampling rate of 1 GSa/s. Thus, one can estimate the conditional min-entropy of the raw data $H_{\min}(M_{dis} | E) = 11.0407$ [25]. Correspondingly, a Toeplitz hash matrix with the dimension $7168 \times 5120$ integrated in FPGA is developed as the RE to generate the final random bits, achieving a real-time generation rate of 10 Gbps, which is significantly boosted compared with previous similar works [38, 39]. The final random bit sequences have passed all the NIST-STS statistical tests.

***Detail analysis of the DSP stage and randomness extraction for the digital filtering method.*** Detail analysis for the DSP stage in the experiment of the digital filtering method in Fig. 5(b) is as follows. Firstly, power spectrum of the acquired data is calculated, as illustrated in Fig. 7 (a). Benefiting from the high-speed PD and DSO, it is shown that the effective range of the spectrum reaches up to 25 GHz and thus much more frequency band that contains plentiful quantum randomness can be potentially exploited, which provides precious resources for random number generation. For the spectrum of the raw data, a peak near 1.9 GHz that corresponds to the beating frequency is observed, which is unwanted and should be filtered. Especially, from Fig. 7 (a), it is shown that within the range of 4~24 GHz, the power levels of the raw data and the electrical noise are stably around -153 dB and -164 dB, respectively, which indicates that a clearance over 10 dB with bright flatness can be achieved in this range and therefore the frequency band of 4~24 GHz is thus determined as the flat area.

Then a 10-th order elliptic band-pass digital filter (DF) is designed and applied on the acquired data to extract the frequency band of the flat area. Correspondingly, power spectrum of the data after DF is also calculated, as shown in Fig. 7 (b). On one hand, it is observed that the original spike has been attenuated to the power level approximates to that of the electrical noise and thus the periodicity should be in principle eliminated. On the other hand, the frequency band of the flat area has been successfully extracted with the clearance and flatness reserved significantly. Thus, as a result, a bandwidth of the QES up to 20 GHz has been achieved.

After the DF, to lower the autocorrelation between samples, the filtered raw data is down sampled to 40 GSa/s or 20 GSa/s, respectively, and then autocorrelations of the filtered raw data with different sampling rates are calculated, as shown in Fig. 7 (c). It is observed that, when the sampling rate is 80 GSa/s (i.e., without down sampling), the filtered raw data shows the highest autocorrelation. While when the filtered raw data is down sampled to 40 GSa/s, the autocorrelation is comparable but slightly higher than that when down sampled to 20 GSa/s, which are both obviously lower than that without down sampling. Thus, for the balance of the generation rate and randomness, the filtered raw data is down sampled to 40 GSa/s, which is exactly twice the bandwidth of filtered QES [28].

Finally, with the data after the DF, it is calculated that the raw data and electrical noise obey Gaussian distribution with variance $\sigma_M^2 = 2.272 \times 10^{-6} V^2$ and $\sigma_E^2 = 1.511 \times 10^{-7} V^2$, respectively. Thus, given a sampling rage of $2R = 0.02V$ and digitization resolution of $n = 8$ for the DSO, one has $H_{\min}(M_{dis} | E) = 5.546$. Correspondingly, a Toeplitz hash matrix with the dimension $4096 \times 2800$ is employed as the RE to generate the final random bits, achieving an ultra-fast generation rate of 218 Gbps. The final random bit sequences have passed all the NIST-STS statistical tests.

## Acknowledgements

This work was supported in part by the National Key Research and Development Program of China (Grant No. 2020YFA0309704), the National Natural Science Foundation of China (Grant Nos U19A2076, U22A2089, 62101516, 62171418, 62201530), the Sichuan Science and Technology Program (Grant Nos 2023JDRC0017, 2023YFG0143, 2022ZDZX0009 and 2021YJ0313), the Natural Science Foundation of Sichuan Province (Grant Nos 2023NSFSC1387 and 2023NSFSC0449), the Basic Research Program of China (Grant No. JCKY2021210B059), the Equipment Advance Research Field Foundation (Grant No. 315067206).

## Author contributions

J.Y., B.J.X., B.L. and H.G. conceived the idea. B.J.X., B.L. and H.G. supervised the work. J.Y., B.J.X., B.L. and H.G. performed the theoretical analysis and simulation for the physical model. J.Y., M.W., J.Y.D., H.W. and Y.P. conceived and implemented the experiment. J.Y., Y.M.B. and Y.C.Z. acquired the experimental data and performed the data analysis. J.Y.D. and F.F. implemented the randomness extraction integrated in the FPGA under the supervision of Q.S., W.H. and S.Y. M.W. and H.Y.J. validated the NIST randomness tests. J.Y., J.L.L., Y.C.Z. and B.J.X. wrote the manuscript with contributions from all the co-authors.

## Competing interests

The authors declare no competing interests.

## Materials & Correspondence

Correspondence and requests for materials should be addressed to B.J.X. and B.L.